\documentclass[conference,ifacconf]{IEEEtran}
\IEEEoverridecommandlockouts
\usepackage{cite}
\usepackage{amsmath,amssymb,amsfonts}

\usepackage[absolute,overlay]{textpos} % for absolute positioning

\usepackage{algorithmic}
\usepackage{graphicx}
\usepackage{textcomp}
\usepackage{booktabs}
\usepackage{mathtools}
\usepackage{array}
\usepackage{multirow}
\usepackage{xcolor}
\usepackage{listings}

\usepackage{datatool}

\def\BibTeX{{\rm B\kern-.05em{\sc i\kern-.025em b}\kern-.08em
    T\kern-.1667em\lower.7ex\hbox{E}\kern-.125emX}}

\usepackage{scalerel}
\usepackage{tikz}
\usetikzlibrary{svg.path}

\definecolor{orcidlogocol}{HTML}{A6CE39}
\tikzset{
    orcidlogo/.pic={
        \fill[orcidlogocol] svg{M256,128c0,70.7-57.3,128-128,128C57.3,256,0,198.7,0,128C0,57.3,57.3,0,128,0C198.7,0,256,57.3,256,128z};
        \fill[white] svg{M86.3,186.2H70.9V79.1h15.4v48.4V186.2z}
        svg{M108.9,79.1h41.6c39.6,0,57,28.3,57,53.6c0,27.5-21.5,53.6-56.8,53.6h-41.8V79.1z M124.3,172.4h24.5c34.9,0,42.9-26.5,42.9-39.7c0-21.5-13.7-39.7-43.7-39.7h-23.7V172.4z}
        svg{M88.7,56.8c0,5.5-4.5,10.1-10.1,10.1c-5.6,0-10.1-4.6-10.1-10.1c0-5.6,4.5-10.1,10.1-10.1C84.2,46.7,88.7,51.3,88.7,56.8z};
    }
}

\newcommand\orcidicon[1]{\href{https://orcid.org/#1}{\mbox{\scalerel*{
                \begin{tikzpicture}[yscale=-1,transform shape]
                \pic{orcidlogo};
                \end{tikzpicture}
            }{|}}}}

\usepackage{hyperref} %<--- Load after everything else

\usepackage{lipsum}% http://ctan.org/pkg/lipsum
\usepackage{indentfirst}

\usepackage{scrextend}

\usepackage[english]{babel}
\usepackage{blindtext}

\begin{document}

\title{Industry Members' Perceptions about ABET-based Accreditation: An Exploratory Study in a Developing Country}

%\documentclass[conference]{IEEEtran}
%Multiple authors with multiple affiliations:
\author{
    \IEEEauthorblockN{V. Sanchez Padilla\IEEEauthorrefmark{1},\IEEEauthorrefmark{2}\orcidicon{0000-0003-3205-388X},  Albert Espinal\IEEEauthorrefmark{3}\orcidicon{0000-0003-2619-2752}, Jennifer M. Case\IEEEauthorrefmark{1}\orcidicon{0000-0002-0186-9803}, \\Jose Cordova-Garcia\IEEEauthorrefmark{3}\orcidicon{0000-0003-1624-8521}, Homero Murzi\IEEEauthorrefmark{1}\orcidicon{0000-0003-3849-2947}}
    \IEEEauthorblockA{\IEEEauthorrefmark{1}Dept. of Engineering Education, Virginia Polytechnic Institute and State University, Blacksburg, VA 24060, USA
    \\\{vsanchez, jencase, hmurzi\}@vt.edu}
    
    \IEEEauthorblockA{\IEEEauthorrefmark{2}College of Engineering, Universidad ECOTEC, Km 13.5 Samborondón, Samborondón, EC092302, Ecuador}
    \IEEEauthorblockA{\IEEEauthorrefmark{3}Telematics Eng. Program, Escuela Superior Politécnica del Litoral, Guayaquil, EC090112, Ecuador
    \\\{aespinal, jecordov\}@espol.edu.ec}}

%Code for one single affilitation:
%\author{V. Sanchez Padilla \orcidicon{0000-0003-3205-388X},~\textit{Member,~IEEE}, Jennifer M. Case,  Homero Murzi,  \\ Albert Espinal \orcidicon{0000-0003-2619-2752}, Jose Cordova-Garcia \orcidicon{0000-0003-1624-8521} \\
%\IEEEauthorblockA{Escuela Superior Politécnica del Litoral, ESPOL Polytechnic University,\\ Faculty of Electrical and Computer Engineering, P.O. Box 09-01-5863, Guayaquil, Ecuador\\ (wavelasq, rojecoro, ananloay, vladsanc, mfilian)@espol.edu.ec}}

\maketitle
\IEEEoverridecommandlockouts
%\IEEEpubid{\makebox[\columnwidth]{\textbf{979-8-3503-3286-5/23/\$31.00~\copyright2023 IEEE }\hfill} \hspace{\columnsep}\makebox[\columnwidth]{ }}
\IEEEpubidadjcol

%First, add: \usepackage[absolute,overlay]{textpos} % for absolute positioning

% Absolute positioning of copyright notice
\begin{textblock*}{\textwidth}(15mm,5mm) % adjust vertical position here
\centering
\footnotesize © 2024 IEEE. Accepted manuscript version of a paper published in \textit{IEEE Transactions on Education} (2024). The final version and citation suggested are available on IEEE Xplore at \href{https://doi.org/10.1109/TE.2024.3410996}{https://doi.org/10.1109/TE.2024.3410996}
%, including reprinting/republishing for advertising or promotional purposes, creating new collective works, for resale or redistribution to servers or lists, or reuse of any copyrighted components of this work in other works. %The final version is available at: \url{https://doi.org/DOI}

\end{textblock*}

\begin{abstract}
ABET accreditation is an increasingly prominent system of global accreditation of engineering programs, and the assessment requires programs to demonstrate that they meet the needs of the program’s stakeholders, typically industrial potential employers of graduates. To obtain these inputs, programs are required to assemble an advisory committee board. The views of the advisory board on the relevance of the degree outcomes are an essential part of this process. The purpose of this qualitative research study is to explore the viewpoints that industry stakeholders have on  this type of process. The context for the study was an Ecuadorian engineering program which had successfully achieved the ABET accreditation. The study drew on interviews undertaken with industry members who were part of the advisory board. This study focuses on how they perceive the process and the accreditation awarded, analyzing their views of its usefulness, especially in relation to the employability of graduates. Based on the findings, we offer critical insights into this accreditation process when it takes place in contexts beyond highly industrialized countries.

%\textcolor{red}{TRUSTWORTHINESS AND VALIDATION, CRESWELL AND POTH, SLIDES WEEK 12 QUAL - DR. SARAH.

\end{abstract}


% \fi
%\IEEEpeerreviewmaketitle

\section{Introduction} 

Academic programs from higher education institutions (HEI) seek to strengthen their credentials through accreditation, which can be seen as an indicator of quality management and assurance  \cite{Nigsch2013}. At the local level, countries typically have their own regulatory agencies that establish higher education policies and means for checking compliance both at the university and program level \cite{Juanatey2021} . Globally, there are now systems for accreditation of professional programs, notably ABET\footnote{https://www.abet.org/} which accredits post-secondary programs in engineering, computer science, and natural and applied sciences. ABET is aligned with the Washington Accord, a system of mutual recognition of engineering degrees and accreditation, established in 1989 by English-speaking countries such as the United States, the United Kingdom, New Zealand, Canada, and Ireland \cite{Case2017, Patil2007}, but now including 20 signatories from across the globe and not limited to Anglophone countries. 

ABET's assessment includes a focus on educational objectives, learning outcomes, continuous improvement, and curriculum \cite{Alhorani2021}, in which faculty members and industry employers collaborate as stakeholders in advisory committee boards when an HEI pursues ABET accreditation. The assessment of the faculty members involves their workload in teaching and research, academic and professional qualifications, student-faculty ratio, involvement in professional societies, and development in either teaching or non-teaching activities. This evaluation that also involves the industry employers with the institutions identifies the levels of attainments that industry expects from graduates, including feedback on how the HEI can better satisfy industry needs \cite{vitale2020developing}.

Accreditations allow HEIs or academic programs to demonstrate their strength in teaching methodologies \cite{Quilambaqui2019}, assessment processes \cite{McCullough2020} stakeholders' feedback \cite{10.1371/journal.pone.0258807}, curricula enhancement \cite{liu2020implementation}, and learning resources \cite{MenonM2020}. Although the accreditations might comply with standards for national development in specific regions, this is not always the case and aligning national and global standards can prove a challenge for those seeking accreditation from a non-US context  \cite{Mendoza2020, Klassen2020}. 

%Professors and industry employers are part of advisory committee boards as stakeholders when a post-secondary institution pursues accreditation. The assessment of the faculty members involves their workload in teaching and research, academic and professional qualifications, student-faculty ratio, involvement in professional societies, and development in either teaching or non-teaching activities. The evaluation that involves the industry employers with the institutions identifies the agreement attainments, feedback to satisfy industry necessities based on graduates’ skills, or how the institution follows their suggestions \cite{vitale2020developing}. The activities mentioned are not limited and can involve other parameters.

\subsection{Research relevance and contribution}

Engineering programs in industrialized countries align with the "school culture" of academia promoted since the end of the 19th century. This culture is opposed to the shop culture where engineering training originated, and which aims more toward hands-on engineering practice \cite{Case2017, Seely1999}. The school culture focuses on problem-solving orientations based on theoretical mathematics and involving examinations \cite{channell2019establishment}. Moreover, in highly industrialized countries a bachelor’s degree can be complemented by postgraduate professional programs that build further engineering specialization, making the graduate more attractive for industry employment  \cite{Salto2022}.

In developing countries, such as many in Latin America, which are often less engaged in technological innovation, such postgraduate programs are less common and are mainly oriented to fulfill requisites demanded by HEIs \cite{escribano2018desempeno}. %For instance, Latin America does not count on a regional accreditor to evaluate academic programs 
%\cite{petrie2008advancing} according to the higher education reality of their different countries.
This region orients more towards pragmatism \cite{iglesias2006economic}, which can be an engineering paradigm for undertaking system configurations, maintenance, and infrastructure deployment tasks instead of research and development (R\&D). In most cases, R\&D investments are limited or non-existent \cite{starovoytova2017scientific}, being the activities carried out mainly by higher education institutions and subsidized companies \cite{JARRINV2021105602}. Furthermore, the region presents late growth in the postgraduate offerings because of the inequalities in educational background of students and the policies implemented in each country \cite{walker2020tendencias}.

Engineering programs from contexts outside the industrialized countries that are seeking ABET accreditation will thus need to adapt their objectives to meet the accreditor’s criteria, but this accreditation tends to be sought after because of benefits it can bring to the HEI, including student mobility and agreements between institutions to en- tail strong collaborations in research  \cite{green2011lost}.

Based on the framework proposed by Volkwein et al. \cite{Volkwein2004EngineeringCA}, this study seeks to analyze the impact of pursuing global accreditation such as ABET in Latin American countries, given the nature of its industrial base, where the interest of the industrial sector to establish a relationship with academia differs based on the size, technological research or qualified staff that has an HEI \cite{alvarez2019generacion}. There is an understandable concern that global accreditation could lead to a weakening of nationally established teaching practices and approaches. 

In this study, we investigated an academic program that belongs to a college of electrical and computer engineering from a public polytechnic HEI located on the coast of Ecuador. Sanchez Padilla et al. \cite{sanchez2023} presented preliminary findings of this research drawing in two interviews. This extended exploratory study includes themes that emerged with additional analysis from those interviews and three more from other participants, presenting the responses of five industry members, such as employers and practitioners, who have served on the advisory committee board. Two research questions guided this study:

\begin{enumerate}

%  \item What do industry members consider important in terms of the competencies acquired by recent graduates?
%  \item What are industry member's views on the role of international accreditation for the enhancement of engineering programs?

  \item What do industry members that participate in an advisory board committee of an undergraduate engineering program from a developing country consider important in terms of the competencies acquired by recent graduates?
  \item What are industry members' views on the role of international accreditation for the enhancement of an undergraduate engineering program from a developing country?

\end{enumerate}

The present study aims to depict the stakeholders’ viewpoints on global accreditation systems. Our research team focused on the perceptions that come from industry members about the ABET-based accreditation process because of its novelty in the context described. Our anticipation is that this contribution could be of value to other HEIs in similar contexts. who could use these findings to determine the opportunities, advantages, or shortcomings these processes that follow international guidelines could bring to their HEI and their relationship to the industry.

\section{Literature Review}

The research topic of what industry members and employers consider relevant competencies for engineering graduates is not new. Yet, indexed literature is limited about how this phenomenon takes place in developing countries. Similarly, there is a minimum approach to qualitative studies to understand their views and perceptions about international accreditation processes in engineering programs from developing countries. These limitations can restrict part of the literature survey to present but also allow us to strengthen these topics toward inquiring outlooks from industry members that serve on advisory boards, which can lead to reinforcing their participation and liaison to academia.

This section splits into two parts to align with our research questions: the first describes what industry members consider salient in competencies from engineering students and professionals, whereas the second illustrates how industry members collaborate with HEI's academic programs through suggestions or agreements.

\subsection{Salient competencies considered by industry members}

Available research emphasizes the following in relation to learning outcome assessments in the academic context. %through international accrediting bodies, such as ABET (Washington Accord) or ENAEE (European Commission)\footnote{https://www.enaee.eu/}.
Educational methods are continually evolving at HEIs, as well as the industry needs and requirements for engineering graduates to attain current needs in the knowledge of trending topics (industrial revolution 4.0 \cite{RoyandRoy2021}, language programming \cite{asee_peer_41235}, embedded systems and apps development \cite{Collaguazo2020}), or professional skills (communication, teamwork \cite{Garousi2020}, critical thinking, pragmatism, and leadership \cite{Katarina2021}). These industry needs can be used to establish guidelines based on the realities experienced by different societies.

At a regional level, case studies show contextual particularities. Agrawal \& Harrington-Hurd \cite{Agrawal2016} and Pyrhonen et al. \cite{Pyrhonen2019} gathered feedback from postgraduate students and industry representatives to establish recommendations to improve learning outcomes based on what employers seek as generic skills with personal attributes. Khalid \& Qazi \cite{Khalid2015} approached knowledge generation by problem-based practices, indicating that interinstitutional collaborations could benefit stakeholders by adapting different learning techniques and stressing that inputs from industry strengthen academic programs. On the other hand, Fathiyah et al. \cite{Fathiyah2019ConceptualFF} list skills asked by engineering accreditations bodies and compare current skills with the ones demanded by the industry revolution 4.0, highlighting industry as a contributor in the supporting of skill development and provider of skills upgrade to academic institutions.

\subsection{Industry members collaboration with HEIs}

Literature about industry members' perceptions toward accreditation processes focusing on developing countries is also limited. There is research on stakeholder perspectives in industrialized countries that shows how they are recruited to provide suggestions, feedback, or assessment to enhance academic programs \cite{solnosky2022capstone, hirudayaraj2021soft, fagrell2020curriculum}. Industry communities are aware of their responsibility to society, in which the engineering professionals implement innovative and effective solutions that may change conceptions that a society holds beyond the traditional roles of engineers \cite{chan2009global}. Hussain et al. \cite{9274316} remark that the industry contributes to identifying needs that can engage joint research and establish relationships for interaction between faculty and industry members. 

%Industry contribution to HEIs also enhances curricula to establish academic internationalization, making the programs attractive to industrialized countries' models. If developing countries align with them, a direct effect will be to host and engage with international students at their HEIs to promote quality and interculturally student body, gaining academic reputation \cite{altbach2007internationalization}.

Dotong et al. \cite{dotong2015philippine} investigate the significance of local accreditations to achieve transnational recognition from foreign entities, granting an overview of local and international quality assurance mechanisms. This study remarks that besides international accreditations, other components, such as networking, strong relationships with the industry, and internship collaborations involving stakeholders, increase the HEIs' rankings. These authors emphasize that partnerships between industry and academia attract enrolments of foreign student bodies because international accreditations aid the attainment of equivalent standards and qualifications recognized abroad. Similarly, Shafi et al. \cite{Shafi2019} indicate that indirect PEO assessments from industry employers allowed the case study programs to realize regional demands asked by the industry, in which participation and contributions should be planned and agreed upon by all stakeholders before implementing substantial changes.

\section{Conceptual Framework}

\begin{figure}[!htbp]    \includegraphics[width=0.49\textwidth]{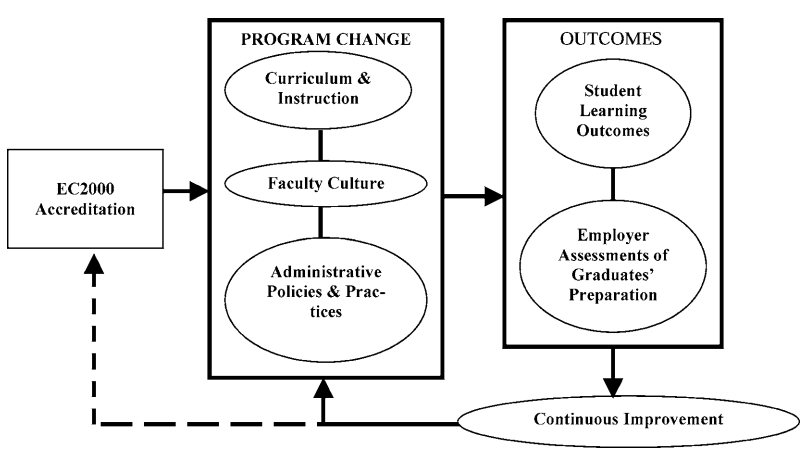}
    \caption{Conceptual Framework proposed by Volkwein et al.}
    \label{fig:Volkwein}
\end{figure}

We chose the conceptual framework proposed by Volkwein et al. \cite{Volkwein2004EngineeringCA} (Figure \ref{fig:Volkwein}), developed to determine if the Engineering Criteria 2000 (EC2000, proposed by the ABET and its stakeholders) impacted student outcomes in engineering programs. This framework postulates that the modified EC2000 accreditation standards should affect curricular changes, instructional methods, assessment initiatives, institutional procedures and reorganization, faculty activities, and values. It was hypothesized that the EC2000 processes and criteria, and the administrative changes resulting from their use, influence student learning outcomes toward an impact on employer assessments of how the students prepare. In addition, through effective constant improvement practices, the information about student learning outcomes and employer satisfaction supplies the foundation for advancements in curriculum and instruction, as well as educational and organizational practices and policy-making.

The framework is relevant to our study as it includes constructs that involve different stakeholders. %For instance, the students’ evaluations in the educational processes require endeavors from faculty members to achieve the objectives traced by the academic programs.
The development of activities might support the revision, modifications, and appraisal of the curriculum and teaching methods implemented. The model proposes that faculty members will lead assessment processes, employ innovative instructional strategies, and boost their awareness of trending topics. %In that sense, it is vital to comprehend their engagement in assessment activities, curriculum reviews, and professional attainment.
For our study, we focused on employers, represented by industry members, in the different roles they can be involved in, such as hiring, supervising, or practicing the profession with the engineer graduates. We considered them relevant because they are implicated in the skill assessments of interns and recent graduates, providing firsthand information on how they perceive their capacities based on the instruction attained by student learning outcomes. They convey feedback to the academic program through the advisory committee boards about the strength the students and graduates require for proper career development outside the educational setting.

From the framework, we focused on the part denoting in employer assessments of graduates’ preparations, which is hypothesized to work in conjunction with the student learning outcomes, feeding into a process of continuous improvement generated from the interaction of these components. Our first research question analyzes employer ratings on the knowledge and professional skills that recent graduates are required to demonstrate in their professional performance. Our second research question focuses more on industry members’ perceptions of the role that accreditation can play on influencing the quality of engineering graduates. %After three years of accreditation. attainment, the responses collected supply an indicator of the effect that industry member stakeholders experience in an ABET-accredited engineering program in a non-US context.

\section{Methods}

This case study is located in Ecuador, classified as a developing country according to the World Economic Situation and Prospects (WEPS) \cite{staff2020world}. %\footnote{https://www.un.org/development/desa/dpad/wp-content/uploads/sites/45/WESP2020\_Annex.pdf}
The context is a polytechnic university with active industry participants collaborating with the technological sector of this country. The purpose of addressing the research questions through a case study is to obtain an in-depth examination of the scenario described \cite{case2011emerging} and to know the perceptions of the participants about the ABET accreditation process through a qualitative methodology based on cross-sectional exploratory research that includes semi-structured interviews in a developing nation context. The following subsections indicate the selection criteria of the participants who collaborated for the qualitative study, the instrument used, the data collection description, and the respective analysis.

%\textcolor{red}{CROSS-SECTIONAL}

%\textcolor{red}{Cite emerging research technologies, Dr. Case}

%\textcolor{purple}{Add: The purpose of addressing the research questions through a case study is to obtain an in-depth examination of the scenario described (Case \& Light, 2011) to know the thoughts and perspectives of the participants about their path to engineering education through a qualitative methodology based on cross-sectional exploratory research that includes semi-structured interviews in a developing nation context (AERA, 2006; Kim, 2021). Before selecting the students, the research team will complete the social and behavioral research training asked by Virginia Tech to seek approval from the Institutional Review Board (Bhattacherjee, 2012; Guillemin \& Gillam, 2004). This study will follow the AERA Council’s suggestions to conduct ethical research involving human subjects, previous determining sampling and perform data  collection (AERA, 2011) to guarantee the principles of voluntary involvement, informed consent, harmlessness, anonymity, and confidentiality, in order the risks posed to the participants are minimal.}

\subsection{Context and participants}

The participants of this exploratory study were industry members that are part of the advisory committee board of an undergraduate engineering program part of the college of electrical and computer engineering from a public polytechnic HEI located on the coast of the Republic of Ecuador that operates with funding from the central government. Besides complying with the national accreditation requirements, the university holds thirteen engineering programs accredited under the ABET accreditation commissions, most of them awarded since 2019, which includes the engineering program we approached. In addition, based on the suggestion of the quality assurance department of the institution, the advisory committee board members are requested to meet annually by the last quarter of the year.  

%\textcolor{red}{Replace Shenton2044StrategiesFE from the below paragraph. Instead, include KRATHWOHL, 2009}

Based on purposive sampling \cite{krathwohl2009sampling}, we drew the selection of the participants according to two criteria, which were that the members had attended the two last board meetings, held in October 2021 and December 2022, and have six or more years of engineering practice relevant to the field of the academic program. The first criterion draws on the continuity in collaboration and attendance at the meeting, and the second because the accreditation process started in 2017. The advisory committee board of 2021 gathered fourteen representatives, while in 2022 was attended by nine. According to the criteria, five industry members met the requirements and agreed to collaborate with the interviews. Table \ref{tbl:01_profiles} depicts their profiles, labeling their identity.

% \usepackage{booktabs}
\begin{table}[]
\caption{Profile of the participants}
\begin{tabular}{@{}ccccc@{}}
\toprule
\begin{tabular}[c]{@{}c@{}}Participant \\ label\end{tabular} & \begin{tabular}[c]{@{}c@{}}Education\\ level\end{tabular} & Gender & \begin{tabular}[c]{@{}c@{}}Industry\\ field\end{tabular} & \begin{tabular}[c]{@{}c@{}}Years of\\ experience\end{tabular} \\ \midrule
Doaldo                                                         & Bachelor's                                                & Male   & R\&D Software                                           & 6+                                                            \\
Jamal                                                         & Master's                                                  & Male   & Data Networks                                            & 15+                                                           \\
Luar                                                         & Master's                                                  & Male   & IP Solutions                                           & 15+                                                           \\
Jena                                                         & Master's                                                  & Female & Cybersecurity                                        & 20+                                                           \\
Ana                                                         & Master's                                                  & Female & Cybersecurity                                        & 10+                                                           \\ \bottomrule
\end{tabular}
\label{tbl:01_profiles}
\end{table}

The second author contacted the participants to ask them about their availability to collaborate with the research. After they agreed, the first and second authors explained the study objective and research purpose, indicating their identities will keep anonymous, passing through a labeling procedure. As the primary language of the participants is Spanish, we set the interviews and followed a coding process in that language to keep the authenticity of the data. In addition, we did not provide any reward for the participants' cooperation to avoid biased situations in their responses.\\

\subsection{Data collection and instrument}

Before selecting the participants, the research team completed the training required to conduct the investigation and obtained informed consent from the Institutional Review Board (IRB) suggested by the AERA Council \cite{council2011aera} to perform the oral interviews. The interviews were our instrument for data collection for bringing in in-depth and insightful responses from the participants \cite{bryman2016social}. We arranged them in person (first option) or virtually (alternatively, but not encouraged). Open-ended responses were invited based on semi-structured interview techniques to procure a framework-informed approach, and we asked follow-up questions when considering something that required further exploration.

%\textcolor{red}{change TIGHT QUALITATIVE for FRAMEWORK INFORMED/DIRECTED}

For the interviews, we employed an interview protocol with three sections defined by topics such as introduction, competencies, and accreditation. It had the flexibility of being modified according to the fluency and depth of the responses throughout the conversations (Table \ref{tbl:02_questions}). We anticipated knowing more about their opinions concerning their awareness of the accreditation process, the perceptions about the graduates, or the enhancements it brings to the integration between academia and industry. We covered the first research question based on the responses collected from the competency queries, whereas we addressed the second research question through the information collected from the queries relevant to accreditation. Besides the interviewees yielding broad responses, they also provided insightful first-hand experience thoughts based on their condition of either hiring recent graduates or dealing with them in junior (entry-level) positions.

\begin{table}[!htbp]
\caption{Interview protocol}
\begin{tabular}{@{}ll@{}}
\toprule
\multicolumn{1}{c}{\textbf{Topics}} & \multicolumn{1}{c}{\textbf{Questions}}                    \\ \midrule
Introduction                      & \begin{tabular}{ m{5.8cm} }
\begin{itemize}
    \item Describe a normal day at work and the challenges you presently face in your industry.
\end{itemize}\end{tabular}

\\ \midrule
Competencies                      & \begin{tabular}{ m{5.8cm} }
\begin{itemize}
    \item At the moment you or your teamwork hire an engineering graduate, do you take into consideration if the graduate comes from an accredited program?
    \item What competencies do you think that engineering graduates should have directly after their graduation?
    \item How do you evaluate the participation of the engineering graduates in the application of engineering principles in their work activities?
    \item If it is applicable to their responsibilities, how do you assess the participation of the engineering graduates in multidisciplinary teams or in activities that demand leadership? \item Do you think the engineering graduates from an accredited program are more competent?
\end{itemize}\end{tabular}

\\ \midrule
Accreditation                            & \begin{tabular}{ m{5.8cm} }
\begin{itemize}
    \item What are your views on the ABET accreditation of engineering programs in Ecuador?
    \item What has been your experience of participating on this Board?    
\end{itemize}\end{tabular} \\ \bottomrule
\end{tabular}
\label{tbl:02_questions}
\end{table}

The interviews were undertaken in the first quarter of 2023, lasting between 25 to 50 minutes in after-hour schedules, including an explanation of the accreditation details to strengthen or confirm what they already know. One interview was conducted in the workplace of the industry member, two at the second author's office, and the last two through video conference. Then, we translated from Spanish to English the thematics found from the transcription and coding process of the interview contents. The thematic analysis focused on the development of candidate themes that resulted from an exhaustive review \cite{ThematicAnalysis}.
%Themes were molded, explained, or rejected 

\subsection{Data analysis and trustworthiness}

Creswell and Poth \cite{creswell2016qualitative} suggest that interviews used for the data analysis strategy %(in our case, onsite or virtually)
be audio recorded, transcribed, and reviewed to attain accuracy. The first and second authors did the interviews and recorded them with the authorization of the interviewees, labeling their identities to guarantee their anonymity. While conducting the interviews, these authors used memoranda to highlight relevant statements from the semi-structured interviews  \cite{KessaRoberts2021}. During the data analysis, the authors did not share the recordings with individuals outside the research team, keeping them in a virtual institutional repository.

The responses from the participants were in Spanish, in which the first author transcribed three of them, and the second did the other two, passing all the information gathered through a cross-check between them to satisfy trustworthiness. For trustworthiness, we sought thick and rich descriptions from the open-ended responses to focus on a thematic analysis \cite{Onwuegbuzie} to provide other stakeholders information for the transferability of the findings \cite{bryman2016social}. We proceeded with the coding process through a web application software for qualitative data analysis using the content from the interviews, memos, and annotations. The responses from the participants were cataloged and summarized through a suitable interpretation process. We worked on coding the patterns for data condensation to approach thematic inductive qualitative analysis \cite{bryman2016social}.

\section{Findings}

%\textcolor{red}{It was also interesting knowing that the interviewees spoke about competencies even though the questions were towards accreditations.}

%\textcolor{red}{Clear overlap what does count as accreditations}

This section depicts the different participants' responses, in which we coded their perceptions and aligned them with the research questions according to themes. For the first research question, a theme emerged from the responses that referred to several competencies that industry members perceive as paramount for recent graduates, and which we determined to be aligned with the ones suggested by ABET throughout an accreditation process. For the second research question, the responses generated two themes with views representing perceptions of the education quality that students received through the following of international standards, and also the lack of awareness by industry members about the process.

\subsection{Competencies that industry require from students and recent graduates}

%Here we align the responses that address RQ1 through the first part of the script: coding related to COMPETENCIES FROM STUDENTS  and COMPETITIVENESS

In this subsection, we illustrate three competencies that industry members and employers agreed as a need for the industry and require for entry-level engineers, which can be applied globally, not differentiating specific economic models or industrial developments.

\subsubsection{Effective communication in workplaces to perform in teamwork}

Concerning establishing effective communication, industry members highlighted this point throughout several interview passages and perspectives, most focusing on how this is essential for teamwork performance. For example, Doaldo highlighted that, based on his experience, he has noticed how recent graduates still struggle to express ideas or even interact with professionals from other fields, often complicating what they wish to communicate by addressing topics using complicated terminology. He recommends that engineers convey ideas using charts or graphics when they result from prototyping, similar to what students learn in design thinking-oriented courses, to achieve effective communication with different audiences. He also pointed out that engineers should communicate in an uncomplicated way their ideas to peers to attain timely production, including documentation with prompts on hardware and software developments, even for work scenarios where teamwork can take place remotely. Jena confirmed the latter due to situations that forced engineering students to communicate remotely, mentioning that:

\begin{addmargin}[1.5em]{0em}
\emph{“Now, they have been studying (indoors), sometimes they get stuck in a lab (practice). They communicate through Teams to follow hints so they can advance with their labs, so they perform well as a teamwork, they are very competent.”}
\end{addmargin}

Other industry members agreed that effective communication leads to better teamwork performance. Jamal indicated that during meetings, recently graduated engineers often do not know how to communicate their ideas to other areas different than engineering, despite all their experience and technical capability. He stressed that it creates setbacks in multidisciplinary teams, as they do not know how to face discussions with the end customers either. On the other hand, Ana agreed that when expressing their ideas, engineers should be empathetic. She provided an example of how an area in charge of network vulnerabilities must get along with staff in charge of development, infrastructure, and information technology audits, underlining that recent engineering graduates should not speak for themselves but also convey a language that allows other peers to understand better, thus avoiding redundancy and time wasting, knowing how to address specific issues to different audiences, determining the communication to carry out between technical staff to administrative or executive officers. She reinforced this point, indicating that knowing how to communicate complements activities related to process management towards leading teams.

\subsubsection{Learning strategies for problem solving}

The study participants gave responses that reflected the importance of problem-solving skills for engineers. For instance, Luar indicated that engineers should aim to brand certifications, especially if they have had solid training to combine theory with pragmatism to address problem-solving. He referred that applying different tools, including math equations, must conduct straightforward applicability to have better action in problem-solving requirements for engineers to set calculations to speed up the solutions asked. Moreover, Jamal mentioned that customers usually request that project engineers have brand certifications to perceive that staff assigned to their accounts are familiar with corporate solutions with vendor support. However, he considers that these strategies should be developed before graduating so that students can opt for certifications that qualify them for the labor market to cope with the technical support challenges that customers demand, including algorithm applications.

Furthermore, Doaldo indicated that academia should encourage students to face challenges that involve complex solutions, primarily if academic programs have international recognition due to accreditation. Although he mentioned that for the hiring process of engineers, employers do not consider whether or not they come from an internationally accredited program, he prefers their profiles to denote necessarily technical knowledge but also the learning desire and to investigate information by themselves, without getting stuck. He considers this a plus that benefits the industry, especially if international accreditation boards promote these skills. A similar response was offered by Jena, indicating that she expects a creative and flexible mentality from engineers, which makes them able to solve complex problems, stating:

\begin{addmargin}[1.5em]{0em}
\emph{“I think that the first competence that engineers, regardless their specialization, should have is being able to solve problems or, better said, to offer creative solutions to complex problems, that is, being able to think outside the box. Thus, I hope that an engineering student who has learned \emph{a} and \emph{b}, will be able to solve \emph{c} (...), showing what (he/she) learned."}
\end{addmargin}

\subsubsection{Task planning}

Even though the participants addressed the topic of functioning effectively as a team in previous lines, their responses also mentioned task planning related to how they can manage their time frames. For example, Jena referred to a specific case in which an engineer, due to her proactivity, sets visits with administrative areas, which she highlighted as positive for integrating with other fields. In addition, Luar commented on an example based on task planning in which engineering staff involved in a project met to propose viable solutions for evaluation and subsequent validation for decision-making with outstanding outcomes. Furthermore, Jamal indicated that task planning and establishing deadlines according to the scheduled activities in teamwork should be paramount to prevent engineering staff from undergoing stress that leads them to performance decline. According to Doaldo, because of the pandemic, engineering students are more familiar with collaboration, task organization, and scope definition tools, all of them for the enhancement of planning activities, mentioning that: \emph{"I realized that a necessary competence is time management, knowing how long a project takes (...) After designing a product that goes to the production stage, engineers have to deal with something new that can show up"}.

These competencies addressed by the industry members align with the student outcomes suggested by ABET, including to know how to communicate effectively in workplaces to perform well in teamwork, to be able to solve complex problems, to be able to engage in continuous learning and to apply the knowledge acquired correctly and, as an additional point, to set goals to optimize timelines for task planning, especially in situations that involve different fields.

\subsection{Education quality through international alignment}

%Here we align the responses that address RQ2 through the second part of the script: coding related to EDUCATION QUALITY, INTERNATIONAL ALIGNMENT, CONTRIBUTION TO THE PROCESS

Regarding the role of an engineering program having international accreditation, participants agreed with points of view indicating that an accreditation process, such as ABET, leads to education quality through alignment. Their opinions were toward improving an education scheme in a combination of theoretical and practical factors aligned with the demands of the labor market. For instance, Luar indicated that enhancing the teaching methodologies for engineering courses strengthens the knowledge received by engineering students to face challenges that the current society demands according to technological evolution, which can result from following suggestions from educational accreditor bodies. Similarly, Jamal indicated that education quality improves competitiveness, demonstrating skills developed in local and international settings, according to the updates required by the industry, which should reflect in the syllabus for students to graduate with appropriate knowledge. He thinks this exposure can help students be competitive in the selection process for a job position.

Jena provided a response that ranked the prestige of a university due to the education quality it can offer. She indicated that she does not consider whether the program is internationally accredited. Instead, she has a professional bias toward engineering professionals who graduate from prestige universities because of the academic quality offered. Her opinion was:

\begin{addmargin}[1.5em]{0em}
\emph{"These international accreditations (...) guarantee that, regardless of which university it is (...) that (academic) program has the quality standards and (...) (necessary) contents, and that students develop their skills and competencies that a professional should have in the field (...), with engineering students who will develop the same skills (...) as the ones from a university from the United States. That is what international accreditations, such as ABET, guarantee to me.”}
\end{addmargin}

Ana, similarly, indicated that international accreditations are significant to demonstrate how an academic program can grow and point towards international standards and not comply only the local ones, concentrating on international guidelines similar to foreign universities, mentioning:

\begin{addmargin}[1.5em]{0em}
\emph{"I would think (it is necessary to have an international accreditation). I honestly think so, because (...) we are trying to equate ourselves (...) to the world standard guidelines and not only the standards asked by the Ecuadorian education law but already looking beyond. And although many people think that certain accreditations or certifications are simply for compliance (...), there are controls and things that must be fulfilled to achieve that role. Those controls, those guidelines make us raise the bar for our (undergraduate) programs. It means, if the program has a certification, I believe that they have quite good global guidelines, and that makes the productive sector improve.”}
\end{addmargin}

Doaldo's criteria focused on indicating that through an accreditation process, education quality assurance is achieved based on standards through rigorous processes, which can reflect how an engineering student or professional from a developing country may have acquired the same knowledge from the \emph{first world}, which he considers can help in many processes related to employment or study abroad based on the suggestions made by ABET towards academic programs that seek to be accredited:

\begin{addmargin}[1.5em]{0em}
\emph{"Processes are something extremely useful, interesting in the sense that the education that I am receiving (...), the series of entities involved in the learning process as such, are the same as in other countries, which is a meaningful fact. (...) as a student and professional, (...) the mindset (…) plays an important role, (knowing that) the education received here is the same as abroad.”}
\end{addmargin}

He complemented his thoughts by indicating the meaning of interacting with others in the workplace, considering this as a plus and a positive influence on the students' and future professionals' attitudes, knowing they were part of a program with international guidelines.

\subsection{Lack of awareness towards the accreditation from employers and industry members}

%Here we address the LACK OF AWARENESS FROM EMPLOYERS AND INDUSTRY MEMBERS coding

Throughout the semi-structured interview, the participants stated that the experience of being part of the advisory board has been gratifying, mostly when the engineering program acknowledged their comments and recommendations for topics to include in the courses or the curriculum modifications to improve the educational process. However, there were opinions from them towards the awareness of the international accreditation of the academic program. Their responses implied that, if they were not members of the advisory board, they would be unaware of this type of process because they have not seen this mentioned in advertisements for both technical and human resources departments, which could be convenient as a decisive point to assess a candidate to occupy a position in the labor market. For instance, Luar varied his answer, indistinctly interchanging the terms that refer to the accreditation of an academic program with the importance of obtaining a brand certification from a specific vendor. He complemented his opinion by indicating that, since there is no adequate process advertisement, there is no appreciation of the corporate sector about the standards followed by academia to have a competitive student. The response given by Jamal was similar, indicating that he has never seen anything about this process or the earned accreditation, either in the press or social networks. Similarly, Doaldo stated that the industry is unaware of the accreditation process and its benefits, even though he realizes that at graduate study applications, universities ask undergraduates if they come from an accredited program.

Further, Ana and Jena, because of their previous collaboration as faculty members of the institution of our case study, are aware of the accreditation process. However, when shifting from academia to industry, similar to the other participants' responses, they said they had not heard about the process or the accreditation earned. Ana indicated that when evaluating candidates for a post, she pays more attention to the institution the prospective engineers come from rather than seeing in their resume a bullet point underscoring they come from an internationally accredited program. She implied that if a vitae emphasizes this, industry members and employers might be interested in learning more about the accreditation, expecting that this diffusion would concern both the academic community and the productive sector. Also, Jena's response indicated that the marketing of this process is perhaps targeted to an audience that she is not a part of, mentioning:

\begin{addmargin}[1.5em]{0em}
\emph{“I would think that perhaps a little more marketing is needed. For example, I know about ABET because I have been a lecturer (...) for many years. I even participated when the first meetings were just taking place (...) to start changing the accreditation programs (...). On the one hand, I am an entrepreneur, but on the other hand, I am a lecturer in a permanent relationship with academia. Then I found out about the certifications, accreditations, etc. But the common businessmen or entrepreneurs, that do not have that link with academia, do not find out that these accreditation processes exist unless someone mentions them, and I have not seen much propaganda about it (…). Maybe they are doing it as direct marketing and that's why I don't hear about it (because) I'm not the target, because I already know, I'm not in that customer database. I would think (...) incurring costs of, for example, advertising could be (...) very expensive, and the university at the moment perhaps does not have the financial resources to do it.”}
\end{addmargin}

Her opinions also focused on how social network postings can support low-budget marketing. She also mentioned more professional networks, such as Linkedin, to create awareness among people from the industry about the importance of the added value of engineering programs accredited by international boards. She thinks engineering programs should communicate to society the benefits these processes bring and the advantages of hiring a professional that was part of these accredited programs.

\section{Discussion}

The results suggest that industry, regardless of the economic model a nation pursues or its income level index classification, tends to require a combination of skills for engineering students or recent graduates aligned to specific recommendations by global accreditor entities, such as the ABET. The works presented in \cite{Charosky2022, Ayadat2020, KnightDaniel2019} confirm this aspect, highlighting the necessity that, before graduates enter the labor market, education must reinforce the development of skills aiming at innovation with the support of infrastructure improvements, laboratories, software acquisition, and teamwork culture. Contrary to our research, those studies included senior students evaluated in initiatives, entrepreneurship, and experiences during capstone projects. For instance, Knight et al. \cite{KnightDaniel2019} focused on real-life practical contexts with a view to a transition to the workplace in communities of practice, a topic addressed by our participants. Despite welcoming these skill developments for technical positions, our findings suggest they have to be considered in non-academic settings. Thus, aligning with a student outcome model may be convenient for engineers to address more general and interdisciplinary issues in different contexts, which generates a plus for their academic degree.

Through the interviews, we can recognize how the industry seeks specific skills for engineers pursuing entry-level positions. The participants agreed that engineering program curricula should emphasize technical and non-technical skills, depicting how the industry requires academic curricular content that aligns with demanded skills and expects their application by engineering graduates \cite{LeandroCruz2022}. This aspect matches the thoughts of some of our participants that engineers have to arrange real-context solutions to be prepared in advance with the labor market, such as effective communication and problem-solving as essential skills asked by industry professionals and employers, such as highlighted in previous literature \cite{WilsonTT2018, Kelley-ASEE2022}.

Based on the responses, we noticed a lack of communication between academia and industry about the engineering programs that align with international accreditations and the benefits they can bring to the labor market by having more competent professionals aligned with international curriculums. Works carried out in industrialized countries suggest the necessity to establish liaisons between local, regional, or international organizations, both academic and productive sectors, by the active participation of stakeholders \cite{Kelley-ASEE2022, derby2020sustainability}. An issue arises when there is no adequate dissemination of the accreditation badges, with possible consequences that both the industry yields highly qualified people for engineering positions and engineers who graduated from rigorous academic programs cannot participate in the constant changes demanded by the productive sectors. Based on these findings, we believe that academia, in the particular case of the developing countries, should work better to advertise the international accreditations attained, suggesting their students and recent graduates promote as well the distinctions that they perceive about the formative process acquired to highlight to the society the differences they have from their peers graduated from non-accredited engineering programs.

Finally, even though other works remark on the need for an enhancement and appraisal of universal competence education and how it can customize toward the industry sector \cite{cordeiro2020production, ortiz2020framework}, our findings further extend the thoughts of our participants on their perceptions that accreditations change education alignments and may even emulate higher education models from industrialized countries, going beyond of the improvement of educational competencies of the academic programs offered by the HEIs. Tacitly, they stressed that the industry needs that engineering students develop skills through learning outcomes suggested by international accreditation bodies. The earning of these accreditations can boost the prestige of an HEI according to the responses obtained, which also advocate that the program achievements should translate to an impact on society in such a way that interaction with academia raises standards and make the labor market more competitive with engineering professionals that correspond to the Sustainable Development Goals determined by the United Nations to approach industry 4.0 \cite{LUPI2022103543}, which can lead to improving economic indicators as well as global academic rankings.

\section{Limitations and Implications for Future Research}

Although we collected in-depth responses from participants with vast industry experience, the criteria for participant selection allowed us to gather only five industry members. The engineering program in which they are part of the advisory board locates in Ecuador. HEIs' academic programs from this country do not have the tradition of pursuing accreditation from foreign organizations. Besides the engineering program mentioned throughout this paper, only two other Ecuadorian HEIs' engineering programs have earned the ABET accreditation, two each. The particularity is that these two institutions are private universities that run with different funding, enrollment conditions, and tuition policies.

Another aspect that shows up as a limitation is that we approached an engineering program from a college of electrical and computer engineering that relies on a curriculum related to data communication, cybersecurity, telemetry, and programming. Its structure differs from other engineering programs focused on earth sciences (e.g., civil engineering), marine sciences (e.g., naval engineering), or production sciences (e.g., systems and industrial engineering), which focus on different educational objectives and curricula. Therefore, our research team is conscious that not all HEIs can have the same management and operations in infrastructure and resources, either for public or private HEIs. 

Current engineering programs from Latin American Spanish-speaking countries awarded with the ABET accreditation include Chile, Colombia, Ecuador, México, and Perú. As not all the countries from the region participate in the ABET accreditation process, we are limited to providing a generalization based on the results of our study. Our research relies on thick description responses for seeking transferability or benchmarking to diverse contexts. Consequently, we suggest that further research include prospective advisory board members, especially ones that lack accreditation awareness to pursue follow-ups of their contributions and thoughts, including other academic programs that follow the suggestions from other accreditor agencies.

Even though our work focused on a case study to know more about industry members' thoughts about accreditation aspects through deep responses, we look forward to contributing insights into the current perception of the industry about the outcomes they notice and perceive from the accreditation process itself, for instance, to enhance research and collaboration with academia or realize tangible/intangible aspects from the recent graduates (or interns) in the industry and to encourage further analysis regardless of the economic standards of the country in which the engineering program develops.

%\textcolor{red}{HOW SOME ANSWERS OVERLAPS}

%\subsubsection{Other aspects: Experience required}

%Here we address the EXPERIENCE FOR A POSITION coding

\section*{Acknowledgment}
The authors would like to thank the industry members of
the advisory committee board for their patience, time, and
contribution to this research.

%Commenting this lines to add the reference block generated:
%\bibliographystyle{IEEEtran}
%\bibliography{biblio.bib}

\vspace{12pt}
\end{document}